\documentclass[11pt]{article}
\usepackage{graphicx}
\usepackage{latexsym}
\usepackage{hyperref}
\def\exp{{\rm exp}\,}

\def\a{\alpha'}

\def\centeron#1#2{{\setbox0=\hbox{#1}\setbox1=\hbox{#2}\ifdim
   \wd1>\wd0\kern.48\wd1\kern-.48\wd0\fi
   \copy0\kern-.48\wd0\kern-.48\wd1\copy1\ifdim\wd0>\wd1
   \kern.48\wd0\kern-.48\wd1\fi}}

\newcommand{\beq}{\begin{equation}}
\newcommand{\eeq}{\end{equation}}
\newcommand{\bea}{\begin{eqnarray}}
\newcommand{\eea}{\end{eqnarray}}
\newcommand{\ba}{\begin{array}}
\newcommand{\ea}{\end{array}}

\newcommand{\p}{\partial}
\newcommand{\nn}{\nonumber}

\renewcommand{\thefootnote}{\fnsymbol{footnote}}
\begin{document}

\hskip3cm

\centerline{\LARGE \bf Hawking radiation as tunneling}
\vskip0.5cm \centerline{\LARGE \bf from charged black holes in 0A string theory}

\vskip2cm

\centerline{\Large Hongbin Kim\footnote{hongbin@yonsei.ac.kr}}

\hskip2cm

\begin{quote}
Department of Physics, College of Science, Yonsei University,
Seoul 120-749, Korea
\end{quote}

\hskip2cm


\vskip2cm

\centerline{\bf Abstract}
There has been much work on explaining Hawking radiation as a quantum tunneling process through horizons.
Basically, this intuitive picture requires the calculation of the imaginary part of the action for outgoing particle.
And two ways are known for achieving this goal: the null-geodesic method and the Hamilton-Jacobi method.
We apply these methods to the charged black holes in 2D dilaton gravity which is originated from the low energy effective theory of type 0A string theory.
We derive the correct Hawking temperature of the black holes including the effect of the back reaction of the radiation,
and obtain the entropy by using the 1st law of black hole thermodynamics.
For fixed-charge ensemble, the 0A black holes are free of phase transition and thermodynamically stable regardless of mass-charge ratio. We show this by interpreting the back reaction term as the inverse of the heat capacity of the black holes. Finally, the possibility of the phase transition in the fixed-potential ensemble is discussed.

\thispagestyle{empty}
\renewcommand{\thefootnote}{\arabic{footnote}}
\setcounter{footnote}{0}
\newpage

\section{Introduction}
Since Hawking proved that black holes can radiate thermally~\cite{Hawking1975}, it is believed that the black holes are kinds of a thermal system and have thermodynamic relations among the quantities describing black holes. There have been especially much efforts to derive the temperature and the entropy of the black holes via various methods. While Hawking used the quantum field theory in curved spacetime in his seminal paper~\cite{Hawking1975}, there exist other methods which give the same predictions. For example, the brick wall method proposed by 't Hooft~\cite{tHooft1985}, or the gravitational anomaly method proposed by Robinson and Wilczek\cite{RobinsonWilczek2005}, or more traditional ways, such as avoiding conical singularity in the Euclidean sector\cite{GibbonsHawking1977}, are suggested. Although all these methods and their modified versions have been successful in deriving the temperature or the entropy of certain types of black holes, they are not satisfactory in the sense that they do not reveal the dynamical nature of the radiation process. The background geometry is fixed mostly in the standard methods. In fact, Hawking and Hartle~\cite{Hartle1976} described the radiation as a pair production. According to this scenario, a pair of particles can be created due to the vacuum fluctuation just inside the horizon. During the positive energy particle tunnel through the horizon to the infinity, the negative energy particle remains inside the hole and reduces the mass of the black hole. However, there is no such intuitive derivation in their original work. As a physical process, it is still important to discuss more realistic models for describing the radiation process.

An idea which treat the Hawking radiation as a semi-classical quantum tunneling process has been proposed by Parikh and Wilczek~\cite{Parikh1999}. Over the past decade, there have been much works applying this method to various types of black holes.(See the references in \cite{Umetsu2011}) The essential idea is that the positive energy particle created just inside the horizon can tunnel through the geometric barrier quantum mechanically. And it is observed as the Hawking flux at infinity. In quantum mechanics, the tunneling probability for classically forbidden paths is given by
\beq
\Gamma \simeq e^{-2\textrm{Im} \textit{I}}, \label{tunneling prob}
\eeq
where $I$ is the action of the particle system.
Though the typical wavelength of the radiation is of the order of the size of the black hole, WKB approximation, or the point particle approximation is justified.\cite{Parikh1999} If we assume that the black holes are in the thermal equilibrium state with the surroundings, the tunneling probability can be identified with the regular Boltzmann factor for a particle of energy $\omega$,
\beq
\Gamma \simeq e^{-\beta \omega},
\eeq
where $\beta$ is the inverse temperature.
One can calculate the Hawking temperature by analyzing the imaginary part of the action for an outgoing particle.
We note that the conservation of energy plays a fundamental role in this approach. The total ADM mass is fixed and the mass of the black hole will decrease while the outgoing particle carries a bunch of energy which amounts to the difference between the total energy and the decreased mass of the black hole.\\
Two different methods have been employed to obtain the imaginary part of the action. One is called the null geodesic method used by Parikh and Wilczek~\cite{Parikh1999}. And the other is called the Hamilton-Jacobi method used by Angheben, Nadalini, Vanzo and Zerbini~\cite{Angheben2005}. In the former method, the contribution to the imaginary part of the action comes from two parts. One is the spacial contribution $\oint p_p d\varphi$, where $p_p$ is the canonical momentum for a particle. The other contribution is a temporal one.\cite{temporal} It is noticed that the tunneling amplitude of the form $\exp(-2\textrm{Im} \int p_p d\varphi)$ is not invariant under the canonical transformations.\cite{Chowdhury2006} The correct expression is $\exp(-\textrm{Im}\oint p_p d\varphi)$ which is identical to (\ref{tunneling prob}) in ordinary quantum mechanics. But they can be, in general, different from each other in general relativity. The temporal contribution must be included because the role of the temporal coordinate and the spacial coordinate are exchanged when we across the horizon. Then it is possible to calculate the imaginary part of the action by using the Hamilton's equation and the metric-dependent information of the null geodesic. The Hamilton-Jacobi method is more direct. One can consider the scalar particles emission and the Klein-Gordon equation governing those particles. Other higher spin particles, of course, can be considered because the Hawking radiation can be composed of particles of various spin. Many authors have succeeded in applying fermion tunneling from various black holes to correctly recover Hawking temperatures~\cite{fermions}. We only consider the neutral scalar particle emission in this letter. By using the WKB approximation, one can obtain the Hamilton-Jacobi equation from the Klein-Gordon equation. Then by solving the Hamilton-Jacobi equation, one can calculate the action directly. Although the original derivation~\cite{Angheben2005} did not consider the self-gravitational effects, a prescription for incorporating the higher order effects of the back reaction is provided.\cite{Medved2005} We also use this modified method to relate the back reaction term to the thermodynamic stability.

In this letter, we apply these two methods to the charged black holes in two dimensional dilaton gravity which is originated from the low energy effective theory of the 0A string theory. The non-chiral projection of two dimensional fermionic string theories gives rise to two types of 0 string theories : type 0A and type 0B. These type 0 theories contain bosonic fields only because NS-NS and RR sectors survive under the non-chiral projection. In type 0A string theory, the NS-NS sector contains a graviton, a dilaton, and a tachyon, while RR sector includes two one-form gauge fields~\cite{Douglas2003}\cite{Takayanagi2003}. It is well known that the low energy effective theory of 0A strings admits charged black hole solutions~\cite{Gukov2003}~\cite{Berkovits2001}. We first derive the correct Hawking temperature via the above two methods and obtain the back reaction term which appears naturally in the tunneling framework. By using the thermodynamic relations, we read off the heat capacity from the back reaction term and show that the heat capacity is positive definite. For usual charged or rotating black holes(for example, Reissner-Nordstr\"{o}m or Kerr-Newman) there exist the critical point at which the second order phase transition occurs.\cite{Davies1977} The heat capacity becomes infinite and discontinuous at that point which characterizes the second order phase transition. And the critical point is determined by the ratio between the the mass, the charge, and the angular momentum. Unlike these black holes, 0A black holes do not allow the negative heat capacity in a physically relevant region. The analysis depends on the ensemble we are dealing with. We mainly use the fixed-charge ensemble in which the thermodynamic stability holds.

The organization of the paper is as follows: In section 2, we briefly review the black hole solutions in type 0A string theory. In section 3, we apply the null geodesic method to the black holes to calculate the imaginary part of the action. After comparing the result with the Boltzmann factor for a particle of energy $\omega$, we can reproduce the Hawking temperature at the leading order in $\omega$. And we obtain the self-interaction effect as a result of applying the energy conservation. We also use the Hamilton-Jacobi method to derive the consistent result. In section 4, we read off the heat capacity from the $\omega^2$ term and the thermodynamic stability is discussed. Section 5 ends with conclusions and discussions on the possibility of the phase transition in the fixed-potential ensemble.

\section{Review of black holes in type 0A string theory}

In this section, we briefly review the black hole solutions in the low energy effective theory of the 0A strings.

The low energy effective action of type 0A string theory at the
lowest order in $\a$ is given by~\cite{Douglas2003}
\bea I_{0A} &=& \int d^2 x \sqrt{-g}\,\bigg[\frac{1}{2\kappa^2}~
e^{-2\Phi}\Big(R + 4\nabla_{\mu}\Phi\nabla^{\mu}\Phi +
\frac{8}{\a} -f_{1}(T)(\nabla T)^2 + f_{2}(T)\Big) \nn \\
&&\qquad \qquad \qquad  -\frac{2\pi\a}{4}f_3(T)(F^{+})^2 -
\frac{2\pi\a}{4}f_3(-T)(F^{-})^2 - Q_{+}F^{+} - Q_{-}F^{-}
\bigg]\,,\label{0AString0}\eea
where $F_{\pm}$ denote field strengths of two RR gauge fields and $Q_{\pm}$ denote the corresponding charges, respectively.
The theory admits the following linear dilaton geometry as a vacuum solution:
\bea \Phi &=& -k\varphi\,, \nn \\
ds^2 &=& -dt^2 + d\varphi^2\,,
\label{vacuum}\eea
where all other fields vanish.
 It is also known that there exist charged black hole solutions in the model~\cite{Gukov2003}\cite{Berkovits2001}.
When the tachyon field $T$ is turned off, RR gauge fields can be
easily solved as
\begin{equation}
F^{+}_{01}=F^{-}_{01}=\frac{Q}{2\pi\a}\,,\qquad  T=0~,
\end{equation}
which corresponds to the configuration of the background D0-branes
with charges given by $Q_{\pm}=Q$\,.

One may use this to integrate out RR gauge fields, and obtain the action of the form
\beq I_{0A}' = \int d^2 x \sqrt{-g}\,\bigg[\frac{1}{2\kappa^2}~
e^{-2\Phi}\Big(R + 4\nabla_{\mu}\Phi\nabla^{\mu}\Phi + 4k^2  \Big) +\Lambda
\bigg]\,.\label{0AString1}\eeq
Here we denoted the original cosmological constant as $k^2 =
2/\alpha'$ and a new effective cosmological constant, coming from
the gauge field contributions, as $\Lambda=-Q^2/(2\pi\a)$.
Therefore the low energy effective theory of type 0A string theory
reduces to the two dimensional dilaton gravity with two kinds of
cosmological constants, one of which is related to the charges of
RR gauge fields. The theory admits the charged black hole
solutions in which the dilaton is taken to be proportional to the
spatial coordinate $\varphi$,
\begin{equation}
\Phi = -k\varphi
\end{equation}
while the black hole geometry is of the form
\beq ds^2 = -f(\varphi) dt^2 + \frac{d\varphi^2}{f(\varphi)} \,,\label{metric}
\eeq
with the factor $f(\varphi)$ given by\footnote{We set $2\kappa^2=1$ for brevity}
\beq f(\varphi) = 1 - \frac{1}{2k} e^{-2k\varphi}
\Big(M-\Lambda\varphi\Big) \,.
\eeq
$M$ may be regarded as a mass of the black hole.
The horizon of the black hole, as an implicit function of $M$, is given by,
\beq
e^{-2k\varphi_H}(M - \Lambda \varphi_H) = 2k. \label{horizon}
\eeq
The extremal black hole solutions are obtained by imposing the following two conditions,
\beq
f(\varphi_{ex})=0, \qquad f'(\varphi_{ex})=0
\eeq
From these conditions one can express the position of the horizon and the mass in terms of the charge,
\beq e^{2k\varphi_{ex}} = - \frac{\Lambda}{4k^2}\,, \qquad M_{ex}
= -\frac{\Lambda}{2k}\Big[1-\ln\Big(-\frac{\Lambda}{4k^2}\Big)\Big]
\,. \eeq
Then the extremal black hole geometry can be expressed as
\beq
f(\varphi) = 1 - e^{-2k(\varphi-\varphi_{ex})}\Big(2k(\varphi-\varphi_{ex}) + 1 \Big).  \label{extremal}
\eeq

\section{Hawking radiation as tunneling}
In this section, we derive the Hawking temperature and the back reaction effect of the 0A black holes on the basis of tunneling mechanism. Two methods are used in calculating the imaginary part of the action : the null geodesic method and the Hamilton-Jacobi method.

\subsection{Null geodesic method}
In order to describe the `across-horizon' process, it is necessary to adopt the Painlev\'{e} coordinates which have no coordinate singularity at the horizon.
Introduce new time coordinate to the metric (\ref{metric}) by
\beq
d t \rightarrow d t - \frac{1}{f(\varphi)}\sqrt{1-f(\varphi)}d\varphi. \label{Painleve}
\eeq
Then the line element becomes
\beq
ds^2 = - f(\varphi)dt^2 + 2\sqrt{1-f(\varphi)}dtd\varphi + d\varphi^2. \label{metric2}
\eeq
The null geodesic for the metric (\ref{metric2}) is given by
\bea
\dot{\varphi} &=& \pm 1-\sqrt{1-f(\varphi)} \nn \\
&=& \pm1-\frac{1}{\sqrt{2k}}e^{-k\varphi}\sqrt{M-\Lambda \varphi},
\eea
where the upper(lower) sign correspond to the outgoing(ingoing) geodesics. We take the positive sign to consider the outgoing geodesic which is forbidden classically. On the other hand, the ingoing particle can be ignored because the probability amplitude of the ingoing particle is unity so that the net result of the loop integral $\oint p_p d\varphi$ is the same with that of outgoing particle only.\cite{Chowdhury2006} \\
\beq
\exp \Big(-\textrm{Im}\oint p_p d\varphi \Big) = \exp \Big(-\textrm{Im}\int_{\varphi_{in}}^{\varphi_{out}} p_p d\varphi \Big)
\eeq
The imaginary part of the action for an outgoing positive energy particle is given by
\beq
\textrm{Im} \int_{\varphi_{in}}^{\varphi_{out}} p_p d\varphi = \textrm{Im} \int_{\varphi_{in}}^{\varphi_{out}} \int_0^{p_p} d p_p' d\varphi =  - \textrm{Im} \int_{\varphi_{in}}^{\varphi_{out}} \int_M^{M-\omega} \frac{d M'}{\dot{\varphi}(\varphi ; M')}d\varphi,
\eeq
where $p_p$ stands for the canonical momentum of an outgoing particle. In canonical theory, the action actually contains the term involving the integration of the Hamiltonian over time, $d I = p_p d\varphi - H_p dt$.($H_p$ is the Hamiltonian of the particle system.) When we consider the closed loop integral which is invariant under the canonical transformations, the spacial part $\textrm{Im} \int p_p d\varphi$ gives twice the Hawking temperature and the temporal part $H_p \textrm{Im} \Delta t$ gives the exact compensation to yield the well-known value for the Hawking temperature. Essentially the temporal contribution is related to the Painlev\'{e} coordinates (\ref{Painleve}) in which the time coordinate has a pole if one integrates across the horizon. The temporal contribution is a generic feature of the tunneling picture.\cite{temporal} In the last equality above, we have used the Hamilton's equation, $\dot{\varphi}=d H_p / d p_p$ at fixed $\varphi$ and used the energy conservation, $M' = M - \omega'$. Note that the eigenvalue of the Hamiltonian $H_p$ is identified with $\omega$.\\
Then the metric component of the geometry into which the outgoing particles tunnel through is
\beq
f(\varphi) = 1 - \frac{1}{2k} e^{-2k\varphi}
\Big(M-\omega-\Lambda\varphi\Big) \,.
\eeq
The spacial contribution to the imaginary part of the action can be obtained by using Feynman's $i \epsilon$ prescription
\bea
\textrm{Im} \int_{\varphi_{in}}^{\varphi_{out}} p_p d\varphi &=& - \textrm{Im}  \int_{\varphi_{in}}^{\varphi_{out}} \Big[ P \int \frac{d M'}{\dot{\varphi}(\varphi ; M')} + i \pi \int \delta(\dot{\varphi}(\varphi ; M'))d M' \Big] d\varphi   \nn \\
&=& -\pi \int_{\varphi_{in}}^{\varphi_{out}} \int_M^{M-\omega} \delta \Big( 1-\frac{1}{\sqrt{2k}}e^{-k\varphi}\sqrt{M'-\Lambda \varphi} \Big)d M' d\varphi \nn \\
&=& -\pi  \int_{\varphi_{in}}^{\varphi_{out}} 4k e^{2k\varphi} d\varphi \nn \\
&=& \frac{\pi}{k} [\omega - \Lambda(\varphi_{in}-\varphi_{out})], \label{ImS}
\eea
where the sign convention is fixed by requiring the positivity of the coefficient in $\omega$, i.e, inverse temperature at the leading order.\\
To proceed further we have to know the full $\omega$ dependence of $\varphi_{in}-\varphi_{out}$, but this is impossible because of the transcendental character of the metric component,
\beq
e^{-2k\varphi_{in}}
(M-\Lambda\varphi_{in} )=2k \, , \quad e^{-2k\varphi_{out}}
(M-\omega-\Lambda\varphi_{out} )=2k. \label{inout}
\eeq
If we perform a series expansion in $\omega$,
\beq
\varphi_{out} = \varphi_{in}(M-\omega) = \varphi_{in}(M) + \Delta \varphi(\omega),
\eeq
where $\Delta \varphi = \Delta \varphi_1 \omega + \Delta \varphi_2 \omega^2 + \cdots $, we get the results from (\ref{inout}) as follows,
\beq
\Delta \varphi_1 = \frac{-1}{\Lambda + 2k(M-\Lambda \varphi_{in})},
\eeq
\beq
\Delta \varphi_2 = \frac{-2k^2(M-\Lambda \varphi_{in})}{[\Lambda + 2k(M-\Lambda \varphi_{in})]^3}.
\eeq
Then the WKB probability amplitude for the classically forbidden trajectory is given by
\bea
\Gamma &=& \exp \Big[-\textrm{Im} \Big(\oint p_p d\varphi - \omega \Delta t^{in} - \omega \Delta t^{out} \Big)\Big] \nn \\
&=& \exp \Big[
\frac{-4\pi (M-\Lambda \varphi_H)}{\Lambda + 2k(M-\Lambda \varphi_H)}\omega + \frac{2\pi k\Lambda(M-\Lambda\varphi_H)}{[\Lambda + 2k(M-\Lambda \varphi_H)]^3}\omega^2 + \mathcal{O}(\omega^3) \Big]. \label{WKBamp}
\eea
We see that the first term proportional to $\omega$ gives the surface gravity of the black hole.
By comparing with the regular Boltzmann factor for a particle of energy $\omega$, $\Gamma = \exp(-\beta \omega)$, the Hawking temperature is given by
\beq
T_H = \frac{k}{2\pi}\Big[ 1+\frac{\Lambda}{4k^2}e^{-2k\varphi_H} \Big]. \label{temperature}
\eeq
And the quadratic term in $\omega$ is the correction by back reaction of the radiation. This self-interaction effect comes from the energy conservation. Its physical meaning will be discussed in section 4.\\
One can obtain the entropy of the black holes by using standard thermodynamic relations
\beq
d M = T_H d S + \Phi d Q + \Omega d J.
\eeq
There is no angular momentum in the 2D black holes we are concerning about. And we let the charge of the black holes be fixed, therefore we ignore the process of so-called the super radiance. Then the entropy of the black hole becomes
\bea
S &=& \int \frac{dM}{T_H (M)} \nn \\
&=& \frac{2\pi}{k} \int \frac{d \varphi_H}{1+\frac{\Lambda}{4k^2}e^{-2k\varphi_H}}\frac{d M}{d \varphi_H} \nn \\
&=& 4\pi e^{2k\varphi_H}.  \label{entropy}
\eea
This is in agreement with the previous results in \cite{Davis2004} and \cite{Hur2008} which are obtained by using the on-shell Euclidean action formalism and the Wald's Noether charge method respectively.\\

\subsection{Hamilton-Jacobi method}
We now calculate the imaginary part of the action making use of the Hamilton-Jacobi equation for the emissive particles.~\cite{Angheben2005}
When black holes emanate radiation at certain temperature, there can be particles of various spin with the same temperature.
Here we consider a massive scalar field in the background metric (\ref{metric}) satisfying the Klein-Gordon equation.\\
\beq
\hbar^2 \nabla^2 \Phi - m^2 \Phi = 0.
\eeq
By using the WKB approximation,
\beq
\Phi = \exp(-\frac{i}{\hbar}I),
\eeq
one can obtain the relativistic Hamilton-Jacobi equation with the limit $\hbar \rightarrow 0$, which gives the action as a function of the coordinates.
\beq
g^{\mu\nu}\p_\mu I \p_\nu I + m^2 = 0.
\eeq
For the metric (\ref{metric}), the Hamilton-Jacobi equation becomes
\beq
-\frac{(\p_t I)^2}{f(\varphi)} + f(\varphi)(\p_\varphi I)^2 + m^2 = 0.
\eeq
We seek a solution of the form
\beq
I(t, \varphi) = -\omega t + W(\varphi).
\eeq
Solving for $W(\varphi)$ yields
\beq
W(\varphi) = \int \frac{d \varphi}{f(\varphi)}\sqrt{\omega^2 - m^2 f(\varphi)}.
\eeq
It is emphasized~\cite{Angheben2005} that we need to adopt the proper spatial distance,
\beq
d\sigma^2 = \frac{d \varphi^2}{f(\varphi)},
\eeq
which is coordinate invariant.\\
By taking the near horizon approximation
\beq
f(\varphi) = f'(\varphi_H)(\varphi - \varphi_H) + \cdots,
\eeq
we find that
\beq
\sigma = \frac{2\sqrt{\varphi-\varphi_H}}{\sqrt{f'(\varphi_H)}},
\eeq
where $0 < \sigma < \infty$.\\
Then the spatial part of the action function reads
\bea
W(\varphi) &=& \frac{2}{f'(\varphi_H)}\int \frac{d\sigma}{\sigma}\sqrt{\omega^2 - \frac{\sigma^2}{4}m^2 f'(\varphi_H)^2} \nn \\
&=& \frac{2\pi i \omega}{f'(\varphi_H)} + \textrm{real contribution}.
\eea
Therefore we obtain the same result with the null geodesic method and the Euclidean method.
\beq
T_H = \beta^{-1} = \frac{\omega}{\textrm{Im} \textit{I} } = \frac{k}{2\pi}\Big[ 1+\frac{\Lambda}{4k^2}e^{-2k\varphi_H} \Big].
\eeq
Here, we also included the temporal contribution which is exactly the same with $\textrm{Im}W$.
However, the above method does not include the back reaction effect. We may incorporate this higher order effect by replacing,~\cite{Medved2005}
\beq
W=\frac{2\pi i \omega}{f'(\varphi_H)} \rightarrow W_q = \int_0^\omega \frac{2\pi i d \omega'}{f'(\varphi_H(M-\omega'))}, \label{actionfunction}
\eeq
where the subscript $q$ denotes the corrected action. Here a gradual transition from $M$ to $M-\omega$ is considered in order to make this energy decrease compatible with the uncertainty principle.\\
Because the explicit form of $\varphi_H$ in terms of $M-\omega'$ is unknown(see (\ref{horizon})), we use
\beq
W_q = \int_{\varphi_{in}(M)}^{\varphi_{out}(M-\omega)} \frac{2\pi i \,d \varphi_H}{f'(\varphi_H)}\frac{d \omega'}{d \varphi_H}
\eeq
instead of (\ref{actionfunction}). In this way we obtain the same result with (\ref{ImS}) and therefore
\beq
\textrm{Im}I =  \frac{4\pi (M-\Lambda \varphi_H)}{\Lambda + 2k(M-\Lambda \varphi_H)}\omega - \frac{2\pi k\Lambda(M-\Lambda\varphi_H)}{[\Lambda + 2k(M-\Lambda \varphi_H)]^3}\omega^2 + \mathcal{O}(\omega^3).
\eeq

\section{$\omega^2$ term and thermodynamic stability}
In this section, we discuss about the $\omega^2$ term and its relation to the thermodynamic stability.\\
According to \cite{Medved2005}, the corrected action (\ref{actionfunction}) can be written in a Taylor expanded form by,
\beq
W_q = \frac{i}{2}\int_0^\omega \beta(M-\omega')d \omega' = \frac{i}{2}\Big[ \omega \frac{\p S}{\p M} - \frac{\omega^2}{2}\frac{\p^2 S}{\p M^2} + \mathcal{O}(\omega^3)\Big],
\eeq
where we used the 1st law of thermodynamics, $\beta(M) = \frac{\p S}{\p M} = 4\pi/f'(\varphi_H)$.
And the coefficient of the $\omega^2$ term above is related to other thermodynamic quantities in the fixed-charge ensemble as follows,
\bea
\frac{\p^2 S}{\p M^2} &=& \frac{\p}{\p M} \Big( \frac{1}{T(M)} \Big) \nn \\
&=& -\frac{1}{T^2}\frac{\p T}{\p M} \nn \\
&=& -\frac{1}{T^2}\frac{1}{C_Q}
\eea
where $C_Q$ is the heat capacity of the black holes.
\beq
C_Q = - \frac{1}{T^2}\Big( \frac{\p^2 S}{\p M^2} \Big)^{-1}
\eeq
Just by reading off the coefficient of the $\omega^2$ term,
\beq
\frac{\p^2 S}{\p M^2} = \frac{8\pi k \Lambda(M-\Lambda\varphi_H)}{[\Lambda + 2k(M-\Lambda \varphi_H)]^3},
\eeq
we can verify that the heat capacity of 0A black holes are positive definite in a physically relevant region, that is, $T>0$.
\beq
C_Q = -4\pi e^{2k\varphi_H} \Big( 1 + \frac{4k^2}{\Lambda}e^{2k\varphi_H} \Big)  > 0. \label{heat capacity of 0A}
\eeq
For usual charged or rotating black holes, there exist so called the Davies point at which the heat capacity is singular and have different signs in both side around the point. For example, Reissner-Nordstr\"{o}m black holes have a Davies point at $Q=\frac{\sqrt{3}}{2}M$~\cite{Davies1977}, and the second order phase transition occurs at the point. Recently, the physical meaning of the Davies critical point within a tunneling framework was reported focusing on the Reissner-Nordstr\"{o}m and Kerr-Newman black holes.~\cite{La2010} Still it is believed that this critical phenomena are generic for any charged or rotating black holes. But we observe that 0A black holes are quite exceptional : there is no second order phase transition in the fixed-charge ensemble.\\
Let us complete the argument with more rigorous thermodynamic language. In the fixed-charge ensemble, the 1st law of thermodynamics can be written as
\beq
d M_Q = T_Q d S + \Phi d Q,
\eeq
by regarding the parameters $S$(the entropy) and $Q$(the RR charge) as a complete set of global quantities for the black hole thermodynamics. Then the fundamental equation is derived from (\ref{horizon}) and (\ref{entropy}) as follows,
\beq
M_Q(S,Q)=\frac{k S}{2\pi} - \frac{k Q^2}{8\pi}\ln \Big( \frac{S}{4\pi} \Big).
\eeq
And the temperature and the RR potential are given by
\bea
T_Q(S,Q) &=& \frac{\p M_Q}{\p S} = \frac{k}{2\pi} - \frac{k Q^2}{8\pi S}, \nn \\
\Phi (S,Q) &=& \frac{\p M_Q}{\p Q} = -\frac{k Q}{4\pi}\ln \Big( \frac{S}{4\pi} \Big). \label{potential}
\eea
To compute the heat capacity, we use $T_Q$ as a function of $S$ with $Q$ held fixed. Then, we find
\beq
C_Q (S,Q) = T \Big( \frac{\p T_Q}{\p S} \Big)_Q^{-1} = \frac{4 S^2}{Q^2} - S. \label{heat capacity}
\eeq
After inserting $S$ into (\ref{heat capacity}), we verify (\ref{heat capacity of 0A}) in a canonical framework.\\
In general, black holes exhibit the negative heat capacity and its physical meaning is clear. Under the thermodynamic perturbation, the black holes with the negative heat capacity is unstable and therefore can not be regarded as a stable equilibrium state. Hence only a microcanonical ensemble is appropriate. But 0A black holes are free of phase transition and thermodynamically stable. In this connection, the instability region is characterized by a non-physical negative temperature. Hence the 0A black holes can be regarded as a stable equilibrium state within a thermal bath and the canonical ensemble can be used.\\

\section{Conclusion and Discussion}
In this paper we explored the thermodynamic properties of the 0A black holes by using the tunneling picture. First, we applied the null-geodesic method and the Hamilton-Jacobi method to calculate the imaginary part of the action for an outgoing emissive particle. We obtained the correct Hawking temperature and the back reaction effect of the radiation. And we found the heat capacity of the 0A black holes from the back reaction effect which is related to the $\omega^2$ term in the action. The heat capacity of the 0A black holes are found to be positive definite in a physically relevant region. Hence we conclude that the 0A black holes are thermodynamically stable and can be described by a canonical ensemble. And there is no second order phase transition in a fixed-charge ensemble. This is remarkable behavior which does not appear in usual charged or rotating black holes.\\
We only treated the fixed-charge ensemble and showed that there is no phase transition. By changing the representation, we observe another facet of the thermodynamic stability. If we adopt the fixed-potential ensemble, i.e., take the Legendre transformation by using (\ref{potential}),
the mass and the temperature of the 0A black holes are given by
\bea
M_\Phi (S,\Phi) &=& \frac{k S}{2\pi} - \frac{2\pi \Phi^2}{k  \ln(S/4\pi)}, \nn \\
T_\Phi (S,\Phi) &=& \frac{k}{2\pi} - \frac{2\pi \Phi^2}{k S \ln(S/4\pi)^2}.
\eea
From this, the heat capacity is found to be
\beq
C_\Phi (\varphi_H,\Phi) = T_\Phi \Big( \frac{\p T_\Phi}{\p S} \Big)_\Phi^{-1} = \frac{4k\varphi_H e^{2k\varphi_H} (4k^4 {\varphi_H}^2 e^{2k\varphi_H} - \pi \Phi^2)}{\Phi^2 (1 + k\varphi_H)}. \label{heat capacity fix po}
\eeq
The heat capacity shows the infinite discontinuity at $\varphi_H = -1/k$. This signals that there might be a meaningful critical point in this ensemble. Since the temperature is positive definite in the range of $| \Phi | < \frac{2k}{e\sqrt{\pi}}$, this point may then be a Davies point. This phenomena at the critical point probably can be classified as a second order phase transition. It will be interesting to study the critical phenomena in the fixed-potential ensemble.\\

{\bf Acknowledgments}

The author would like to thank S.~Hyun and J.~Baek for helpful discussions and comments. Especially, useful comments from D.~Singleton are gratefully acknowledged. This work is supported by the National Research Foundation of Korea(NRF) grant funded by the Korea government(MEST) with the
grant number 2009-0074518.


\end{document}